\documentclass[preprint,aps,prb]{revtex4}

\usepackage{graphicx}
\usepackage{dcolumn}
\usepackage{bm}
\usepackage{latexsym}

\begin{document}

\title{A proof of Jarzynski's non-equilibrium work theorem for
  dynamical systems that conserve the canonical distribution}
\author{E. Sch\"oll-Paschinger and C. Dellago} \affiliation{Fakult\"at
  f\"ur Physik, Universit\"at Wien, Boltzmanngasse 5, A-1090 Wien,
  Austria} \date{\today}

\begin{abstract}
  We present a derivation of the Jarzynski identity and the Crooks
  fluctuation theorem for systems governed by deterministic dynamics
  that conserves the canonical distribution such as Hamiltonian
  dynamics, Nos$\acute\textrm e$-Hoover dynamics, Nos$\acute\textrm
  e$-Hoover chains and Gaussian isokinetic dynamics.  The proof is
  based on a relation between the heat absorbed by the system during
  the non-equilibrium process and the Jacobian of the phase flow
  generated by the dynamics.
\end{abstract}

\pacs{Valid PACS appear here}
\maketitle


\section{Introduction}

During the past decade our understanding of the thermodynamics of
non-equilibrium systems was considerably extended. Two relations have
been derived which are valid for systems arbitrarily far away from
equilibrium: a family of relations which are collectively called the
fluctuation theorems~\cite{fluctEvans,fluctGallavotti,Adv} and
Jarzynski's non-equilibrium work theorem~\cite{JarPRL}.  The latter
theorem relates the equilibrium free energy difference $\Delta F$ to
the statistics of work $W$ carried out on a system during a
non-equilibrium transformation:
\begin{equation}
  \label{eq:jaridentity}
    \overline{e^{-\beta W}}=e^{-\beta\Delta F}.
\end{equation}
Here the overbar indicates an average over many realizations of the
transformation process which is driven by switching an external
parameter from an initial to a final value.  The maximum work theorem,
which states that the average work performed on the system during such
a transformation is bounded from below by the free energy difference,
i.e. $\overline W\ge\Delta F$, is a direct consequence of the
Jarzynski relation. Besides providing new insight into the statistical
mechanics of non-equilibrium processes, the Jarzynski identity has
also been applied to extract equilibrium free energy differences from
computer simulations~\cite{JarPRE,Hummersim,Oberhofer,Lechner,Kofke}
and experiments~\cite{Bustamante,Ciliberto,Wang}.  In particular,
Liphardt {\it et al.} \cite{Bustamante}Ê have determined the free
energy required to unfold a single RNA chain from non-equilibrium work
measurements and Douarche {\it et al.} \cite{Ciliberto} have verified
the Jarzynski equality for a mechanical oscillator that is driven out
of equilibrium by an external force. Another experimental test of the
Jarzynski equality was provided by Wang {\em et al.} \cite{Wang} who
considered a colloidal particle pulled through liquid water by an
optical trap \cite{Reid}.

Jarzynski's theorem has been derived by various means.  While the
original proof~\cite{JarPRL} applies to systems evolving
deterministically with and without coupling to a heat bath, in
subsequent work~\cite{JarPRE} the proof was generalized to stochastic
systems governed by a master equation. The latter approach comprised
systems evolving under Hamiltonian dynamics, Langevin dynamics,
isothermal dynamics and Monte Carlo dynamics.  Alternative proofs for
Markovian stochastic dynamics that satisfies detailed balance were
given by Crooks~\cite{Crooks} and later by Lechner {\it et
  al.}~\cite{Lechner}.  Another elegant derivation of the
non-equilibrium work relation based on the Feynman-Kac theorem is due
to Hummer and Szabo~\cite{Hummer}, who also discussed how one can
reconstruct free energies from non-equilibrium single-molecule pulling
experiments. Crooks~\cite{CrooksPRE} also derived a generalized
version of a transient fluctuation theorem~\cite{Adv} - in the
following called the Crooks fluctuation theorem - for Markovian
stochastic dynamics that is microscopically reversible. He showed in
subsequent work~\cite{CrooksPRE2} that the Crooks fluctuation theorem
and the Jarzynski equality can be regarded as special cases of a
single theorem that follows if the system is microscopically
reversible. This condition holds for Markovian dynamics that preserves
the equilibrium distribution if left unperturbed~\cite{CrooksPRE2}.
Jarzynski~\cite{Jarjstatp} and Evans~\cite{Evans} derived the Crooks
fluctuation theorem also for deterministic, thermostatted systems.
The transient fluctuation theorems~\cite{Adv} and Jarzynski's
non-equilibrium work relation~\cite{JarPRL} are closely related, as
highlighted in a recent paper~\cite{Reid}.

In this paper we present an alternative proof of the Jarzynski
identity under rather general assumptions: we consider a system
evolving according to deterministic equations of motion that conserve
the canonical distribution. The system, which is initially
equilibrated at a given temperature, can either be decoupled from the
heat bath during the switching process or remains coupled to the heat
bath. In the latter case the influence of the heat bath is described
by introducing one or more variables in an extended phase space and
the time evolution of the system is non-Hamiltonian with equations of
motion that do not conserve phase space volume.  In contrast to prior
derivations, we highlight in our proof the relation between the change
in the phase space volume and the heat absorbed by the system during
the non-equilibrium process.  We show that this relation is strictly
valid for systems that conserve the canonical distribution when the
external control parameter is held fixed.  Various dynamical systems
obey the condition mentioned above, i.e.  they leave the canonical
distribution invariant under the time evolution: this is the case for
Nos$\acute\textrm e$-Hoover dynamics, Nos$\acute\textrm e$-Hoover
chains, Gaussian isokinetic dynamics and Hamiltonian dynamics and the
proof presented here is valid for all these systems. A recent
derivation of the Jarzynski identity by Cuendet~\cite{Cuendet} can be
regarded as a special case of our proof.

The paper is organized as follows. The proof of the Jarzynski identity
is the subject of section II. In section III we derive the Crooks
fluctuation theorem using the same formalism. In section IV we discuss
a few examples of dynamical systems for which the proof is valid.

\section{Jarzynski's theorem}
\label{proof}
Consider a classical system with Hamiltonian $H(x,\lambda)$ depending
on the internal variables $x$ (such as the positions and momenta of
some particles in a box or the magnetic moments of a spin system) and
an external parameter $\lambda$ (such as the volume of the box
confining the particles or an external magnetic field applied to the
spin system).  For a given value of the external parameter $\lambda$
the canonical partition function is given by
\begin{equation} \label{eq:partition_function} 
Z_\lambda=\int \mathrm{d}x \ e^{-\beta H(x,\lambda)},
\end{equation}
where $\beta=1/k_BT$, $T$ is the temperature and $k_B$ the Boltzmann
constant. The Helmholtz free energy of the system is then given by
\begin{equation} \label{eq:Helmholtz} 
F_\lambda=-k_BT\ln Z_\lambda.
\end{equation}
Here we have neglected the prefactor of the partition function which
is irrelevant for the present discussion.

We now consider a process where the external parameter $\lambda$ is
switched in a finite amount of time $\tau$ from some initial value
$\lambda_A$ to some final value $\lambda_B$ starting from a system
that is initially equilibrated with a heat bath of temperature $T$.
The time dependence of $\lambda$, $\{\lambda(t),0\le t\le \tau\}$, is
completely arbitrary and in general the system will be driven out of
equilibrium during the process.  For a given realization of the
process the time evolution of the system is given by the trajectory
$\{x(t),0\le t\le \tau\}$ that can be either deterministic or
stochastic and determines the amount of work $W$ that is performed on
the system during that realization. We repeat this process infinitely
many times where the external parameter is always varied in the same
manner and consider the ensemble of such non-equilibrium processes.
Jarzynski has shown that the average of the work exponential
$e^{-\beta W}$ in the ensemble of non-equilibrium processes can be
related to the free energy difference between the equilibrium states
corresponding to the initial and final values of the external
parameter, $\Delta F=F_B-F_A$, via
\begin{equation}
  \label{eq:jartheorem}
  \overline{e^{-\beta W}}=e^{-\beta\Delta F}.
\end{equation}
Here the overbar denotes an average over all realizations of the
non-equilibrium process starting from canonically distributed initial
conditions. Equation~(\ref{eq:jartheorem}) is Jarzynski's
non-equilibrium work theorem.  In the following, we will prove this
relation for a system evolving deterministically in time.  That is
either the case for Hamiltonian dynamics where the system is decoupled
from the heat bath during the switching process or for non-Hamiltonian
isothermal molecular dynamics like Nos$\acute\textrm e$-Hoover
dynamics, Nos$\acute\textrm e$-Hoover chains or Gaussian isokinetic
dynamics where the system remains coupled to the heat bath. In the
latter cases the coupling of the heat bath is purely through the
equations of motion and there is no interaction energy between system
and bath.  The deterministic evolution in the phase space of the
system and bath is given through some equations of motion
\begin{equation}
  \label{eq:eom}
  \dot{\Gamma}=G(\Gamma,\lambda),
\end{equation}
where $\Gamma=\{x,y\}$ includes the variables $x$ describing the
system and possibly some variables $y$ describing the bath.  The
configuration $\Gamma_0$ at time $t=0$ determines uniquely the
configuration $\Gamma_{t}$ a time $t$ later generating an invertible
and differentiable mapping $\phi_{t}$ that describes the phase flow,
\begin{equation} \label{eq:map} 
\Gamma_{t}=\phi_{t}(\Gamma_0).
\end{equation}
For simplicity we denote $\phi_t$ by $\phi$ in the case $t=\tau$. We
assume that the (normalized) equilibrium distribution $g(y)$ of the
bath variables does not depend on the external parameter $\lambda$ and
we define the function $\psi(y)=-k_BT\ln g(y)$.  Thus, we can construct
an extended Hamiltonian as the sum of the total energy of the system,
$H(x,\lambda)$ and $\psi(y)$:
\begin{equation}
  \label{eq:extendedH}
  {\cal H}(\Gamma,\lambda)\equiv H(x,\lambda)+\psi(y).
\end{equation}
The equilibrium distribution of the combined system for a particular value of $\lambda$
then is
\begin{equation}
  \label{eq:equilibrium}
  \frac{e^{-\beta{\cal H}(\Gamma,\lambda)}}{\int\mathrm{d}\Gamma e^{-\beta{\cal H}(\Gamma,\lambda)}}=\frac{e^{-\beta H(x,\lambda)}}{Z_\lambda}e^{-\beta\psi(y)},
\end{equation}
where we have used the fact that the distribution $e^{-\beta\psi(y)}$
is normalized in the $y$-subspace. For a switch of the parameter
$\lambda$ from $\lambda_A$ to $\lambda_B$ the free energy difference
$\Delta F=F_B-F_A$ is given by
\begin{eqnarray} \label{eq:deltaF1}
  e^{-\beta\Delta F}=\frac{Z_B}{Z_A}&=&\frac{1}{Z_A}\int \mathrm{d}x_\tau e^{-\beta H(x_\tau,\lambda_B)}\nonumber\\
  &=&\frac{1}{Z_A}\int \mathrm{d}\Gamma_\tau e^{-\beta {\cal H}(\Gamma_\tau,\lambda_B)}\nonumber\\
  &=&\frac{1}{Z_A}\int
  d\Gamma_\tau e^{-\beta\left[{\cal H}(\Gamma_\tau,\lambda_B)-{\cal H}(\phi^{-1}(\Gamma_\tau),\lambda_A)\right]}e^{-\beta
    {\cal H}(\phi^{-1}(\Gamma_\tau),\lambda_A)},
\end{eqnarray}
where in the first step we have switched from the Hamiltonian of the
subsystem to the extended Hamiltonian and in the second step we have
multiplied and divided the integrand by $e^{-\beta {\cal
    H}(\phi^{-1}(\Gamma_\tau),\lambda_A)}$.  Changing the integration
variables from $\Gamma_\tau$ to $\Gamma_0=\phi^{-1}(\Gamma_\tau)$ yields
\begin{eqnarray} \label{eq:deltaF2}
  e^{-\beta\Delta F}&=&\frac{1}{Z_A}\int \mathrm{d}\Gamma_0 e^{-\beta\left[{\cal H}(\phi(\Gamma_0),\lambda_B)-{\cal H}(\Gamma_0,\lambda_A)\right]}e^{-\beta {\cal H}(\Gamma_0,\lambda_A)}\left|\frac{\partial \phi}{\partial \Gamma_0}\right|\nonumber\\
  &=&\frac{1}{Z_A}\int \mathrm{d}\Gamma_0 e^{-\beta
    {\cal H}(\Gamma_0,\lambda_A)}e^{-\beta\left[{\cal H}(\phi(\Gamma_0),\lambda_B)-{\cal H}(\Gamma_0,\lambda_A)\right]-k_BT\ln\left|\frac{\partial
      \phi}{\partial \Gamma_0}\right|}
\end{eqnarray}
where we have introduced the Jacobian
$\left|\frac{\partial \phi}{\partial \Gamma_0}\right|$ of the mapping
$\phi$ which is also called the phase space compression factor. If we
define a `work function' $W_\phi$~\cite{Lechner,Evans,JarPRE2} as
\begin{equation} \label{eq:workfunction} 
W_\phi={\cal H}(\Gamma_\tau,\lambda_B)-{\cal H}(\Gamma_0,\lambda_A)-k_BT\ln\left|\frac{\partial \phi}{\partial \Gamma_0}\right|
\end{equation}
then Eq. (\ref{eq:deltaF2}) can be rewritten as
\begin{equation} \label{eq:deltaF3} e^{-\beta\Delta
    F}=\frac{1}{Z_A}\int \mathrm{d}\Gamma_0 e^{-\beta {\cal H}(\Gamma_0,\lambda_A)}e^{-\beta
    W_\phi}=\overline{e^{-\beta W_\phi}},
\end{equation}
where the overbar denotes an average over the distribution of initial
conditions, i.e. a canonical distribution in the $x$-subspace, and the
distribution $e^{-\beta\psi(y_0)}$ in the $y$-subspace.
Equation~(\ref{eq:deltaF3}) looks already quite similar to the
Jarzynski identity~(\ref{eq:jartheorem}), but what remains is to show
that $W_\phi$ is indeed equal to the work $W$.  The last term of the
work function is the Jacobian determinant, while the other terms
describe the change of the extended Hamiltonian along the trajectory
starting in $\Gamma_0$.  Using Def.~(\ref{eq:extendedH}) the work
function $W_\phi$ can be written as:
\begin{equation} \label{eq:deltacalH} 
W_\phi=\Delta H+\psi(y_\tau)-\psi(y_0)-k_BT\ln\left|\frac{\partial \phi}{\partial \Gamma_0}\right|,
\end{equation}where
\begin{equation} \label{eq:deltaH} 
\Delta H=H(x_\tau,\lambda_B)-H(x_0,\lambda_A)=Q+W 
\end{equation}
is the change of the total energy of the subsystem during the
switching process. In the last part of Eq.~(\ref{eq:deltaH}) we have
indicated that the energy $H(x,\lambda)$ can change due to two
different reasons. One contribution arises from changes in $x$:
\begin{equation} \label{eq:Q} 
Q=\int_0^{\tau} \mathrm{d}t\nabla_xH(x(t),\lambda(t))\cdot\dot{x}(t).
\end{equation}
$Q$ is the heat transferred from the heat bath to the system during
the switching process. The other contribution originates from the
change in $\lambda$:
\begin{equation} \label{eq:W} 
  W=\int_0^{\tau}\mathrm{d}t\frac{\partial
    H(x(t),\lambda(t))}{\partial \lambda}\dot{\lambda}(t).
\end{equation}
$W$ is the work performed on the system by changing the external parameter.

In the following we will show that if the distribution
$\rho(x,y)=e^{-\beta\left[H(x,\lambda)+\psi(y)\right]}/Z_\lambda$ is a stationary
solution of the Liouville equation
\begin{equation} \label{eq:Liouville} 
\frac{\partial \rho}{\partial t}+\nabla_\Gamma\cdot(\rho \dot{\Gamma})=0
\end{equation} 
for fixed $\lambda$, then the work function $W_\phi$ equals the
physical work $W$ and thus Jarzynski's theorem is valid.

If a phase space density $\rho$ is a stationary solution of the
continuity equation~(\ref{eq:Liouville}), (i.e.,
$\frac{\partial\rho}{\partial t}$=0), it must obey $\nabla_\Gamma\cdot(\rho
\dot{\Gamma})=0$. In the case of $\rho=
e^{-\beta(H(x,\lambda)+\psi(y))}/Z_\lambda$ this reduces to
\begin{eqnarray} \label{eq:dqdt} 
&&\nabla_\Gamma\cdot\left(e^{-\beta (H+\psi)}\dot{\Gamma}\right)=0,
\end{eqnarray}or
\begin{eqnarray} \label{eq:dqdta} 
&&\nabla_x H \cdot \dot{x}+\nabla_y\psi\cdot\dot y-k_BT\nabla_\Gamma\cdot \dot{\Gamma}=0.
\end{eqnarray}
Using Euler's expansion formula~\cite{Euler} for the time derivative
of the Jacobian,
\begin{eqnarray} \label{eq:dqdt1} 
\nabla_\Gamma \cdot\dot{\Gamma}=\frac{d}{dt}\ln\left|\frac{\partial\phi_{t}}{\partial \Gamma_0}\right|,
\end{eqnarray}
we finally obtain
\begin{eqnarray} \label{eq:dqdt2} 
\nabla_x H \cdot \dot{x}+\nabla_y\psi\cdot\dot y-k_BT\frac{d}{dt}\ln\left|\frac{\partial\phi{_{t}}}{\partial \Gamma_0}\right|&=&0.
\end{eqnarray}
The first term of the above equation is just the time derivative of
the heat $\frac{dQ}{dt}$. The second term is the rate of change of the
function $\psi(y)$:
\begin{equation} 
\frac{d\psi}{dt}=\nabla_y\psi(y(t))\cdot\dot y(t). 
\end{equation} 
Integrating
Eq.~(\ref{eq:dqdt2}) from 0 to $\tau$ yields
\begin{equation}
  \label{eq:heateq}
 Q+\psi(y_\tau)-\psi(y_0)-k_BT\ln\left|\frac{\partial\phi}{\partial \Gamma_0}\right|=0,
\end{equation}
which expresses the fact that the heat transfer to the extended
system, $\int_0^\tau dt\nabla_\Gamma{\cal H}\cdot\dot{\Gamma}(t)$,
exactly cancels the term including the phase space compression factor.
Inserting this result into Eq.~(\ref{eq:deltacalH}) we obtain
$W=W_\phi$. With Eq.~(\ref{eq:deltaF3}) this completes the proof of
Jarzynski's theorem:
\begin{equation}
  \label{eq:jarzynsikidentity}
  e^{-\beta\Delta F}=\overline{e^{-\beta W_\phi}}=\overline{e^{-\beta W}}.
\end{equation}
We stress, that the physical work~(\ref{eq:W}) depends only on the
trajectory of the subsystem, however in
Eq.~(\ref{eq:jarzynsikidentity}) we have to average the work
exponential $e^{-\beta W}$ over all degrees of freedom of the extended
system, since the time evolution of the subsystem depends on the
configuration of the extended system at $t=0$.

\section{Crooks Fluctuation theorem}
With some additional assumptions the formalism presented in
Sec.~\ref{proof} can also be used to derive the Crooks fluctuation
theorem~\cite{CrooksPRE,Evans} for dynamics conserving the canonical
distribution.  The Crooks theorem relates the probability that for the
forward process the work $W$ takes a value $C$ to the probability that
along the reverse process it takes a value $-C$.  To derive this
expression we note that for each forward path
$\{(\Gamma(t),\lambda(t)),0<t<\tau\}$ we have a reverse path
$\{(\Gamma^T(\tau-t),\lambda^T(\tau-t)),0<t<\tau\}$ where the
superscript $T$ indicates that quantities that are odd under time
reversal (such as momenta) have changed their sign.  The work, heat,
and energy difference along the reverse path are defined as above for
the forward direction and take the negative of the respective forward
values.

We assume in the following that the extended Hamiltonian $\cal H$ is
invariant under time reversal and that both the forward and the
reverse path start from equilibrium distributions, i.e.
$\rho_A(\Gamma_0)=e^{\beta F_A}e^{-\beta{\cal H}_A(\Gamma_0)}$ and
$\rho_B(\Gamma_\tau^T)=e^{\beta F_B}e^{-\beta{\cal
    H}_B(\Gamma_\tau^T)}$. Then
\begin{equation}
  \label{eq:FT}
  \frac{P_F(W=C)}{P_R(W=-C)}=e^{-\beta\Delta F}\frac{\int \mathrm{d}\Gamma_0 e^{-\beta{\cal H}(\Gamma_0,\lambda_A)}\delta(W_F-C)}
  {\int \mathrm{d}\Gamma_\tau^T e^{-\beta{\cal H}(\Gamma_\tau^T,\lambda_B)}\delta(W_R+C)}.
\end{equation}
Since the Jacobian of the time reversal mapping is unity, i.e.
$\left|\frac{\partial\Gamma^T_\tau}{\partial\Gamma_\tau}\right|=1$,
and the extended Hamiltonian is invariant under time reversal we
obtain
\begin{equation}
  \label{eq:FT2}
  \frac{P_F(W=C)}{P_R(W=-C)}=e^{-\beta\Delta F}\frac{\int \mathrm{d}\Gamma_0 e^{-\beta{\cal H}(\Gamma_0,\lambda_A)}\delta(W_F-C)}
  {\int \mathrm{d}\Gamma_\tau e^{-\beta{\cal H}(\Gamma_\tau,\lambda_B)}\delta(W_R+C)}.
\end{equation}
Inserting the relation 
\begin{eqnarray}
  \label{eq:relation}
  {\cal
    H}(\Gamma_\tau,\lambda_B)&=&H(x_\tau,\lambda_B)+\psi(y_\tau)=H(x_0,\lambda_A)+Q+W_F+\psi(y_\tau)\nonumber\nonumber\\
  &=&H(x_0,\lambda_A)+k_BT\ln\left|\frac{\partial\phi}{\partial\Gamma_0}\right|+\psi(y_0)-\psi(y_\tau)+W_F+\psi(y_\tau)\nonumber\\
  &=&{\cal H}(\Gamma_0,\lambda_A)+W_F+k_BT\ln\left|\frac{\partial\phi}{\partial\Gamma_0}\right|
\end{eqnarray} 
into Eq.~(\ref{eq:FT2}) yields
\begin{equation}
  \label{eq:FT3}
  \frac{P_F(W=C)}{P_R(W=-C)}=e^{-\beta\Delta F}\frac{\int \mathrm{d}\Gamma_0 e^{-\beta{\cal H}(\Gamma_0,\lambda_A)}\delta(W_F-C)}
  {\int \mathrm{d}\Gamma_\tau e^{-\beta{\cal H}(\phi^{-1}(\Gamma_\tau),\lambda_A)}e^{-\beta W_F}\left|\frac{\partial\phi^{-1}}{\partial\Gamma_\tau}\right|\delta(W_F-C)},
\end{equation}
where we have used the fact that the work is odd under time reversal,
i.e. $W_R=-W_F$.  Changing the integration variables in the
denominator from $\Gamma_\tau$ to $\Gamma_0=\phi^{-1}(\Gamma_\tau)$
one obtains
\begin{eqnarray}
  \label{eq:FT4}
  \frac{P_F(W=C)}{P_R(W=-C)}&=&e^{-\beta\Delta F}\frac{\int \mathrm{d}\Gamma_0 e^{-\beta{\cal H}(\Gamma_0,\lambda_A)}\delta(W_F-C)}
  {\int \mathrm{d}\Gamma_0 e^{-\beta{\cal H}(\Gamma_0,\lambda_A)}e^{-\beta W_F}\delta(W_F-C)}\nonumber\nonumber\\
  &=&e^{-\beta\Delta F}e^{\beta C}\frac{\int \mathrm{d}\Gamma_0 e^{-\beta{\cal H}(\Gamma_0,\lambda_A)}\delta(W_F-C)}
  {\int \mathrm{d}\Gamma_0 e^{-\beta{\cal H}(\Gamma_0,\lambda_A)}\delta(W_F-C)}\nonumber\\
 &=&e^{-\beta\Delta F}e^{\beta C}.
\end{eqnarray}
Thus, the Crooks fluctuation theorem is valid provided the dynamics
conserves the extended canonical distribution and the extended
Hamiltonian is invariant under time reversal.

\section{Discussion}
We have shown that the Jarzynski identity is valid for deterministic
dynamics that conserves an extended canonical distribution of the form
of Eq.~(\ref{eq:equilibrium}). Let us now consider a few examples of
such dynamical systems.

\subsection{Hamiltonian dynamics}

For Hamiltonian dynamics the proof of Section II simplifies
considerably since we do not need to consider the $y$-subspace.
Furthermore, the phase space volume is conserved, i.e.
$\left|\frac{\partial\phi}{\partial x_0}\right|=1$, and no heat $Q$ is
absorbed by the isolated system.  Thus the work function $W_\phi$
trivially equals the physical work $W$ and the Jarzynski relation
holds.

One can also consider the case when the system of interest and the
heat bath form a large Hamiltonian system and the coupling between
them is weak. In this case, $\psi(y)$ is just the Hamiltonian of the
heat bath and the phase space compression factor in
Eq.~(\ref{eq:heateq}) is identically zero. So, Eq.~(\ref{eq:heateq})
states that the heat $Q$ absorbed by the system is just equal to the
negative change of the internal energy of the heat bath,
$Q=-\left\{\psi(y_\tau)-\psi(y_0)\right\}$.

\subsection{Nos$\acute\textrm e$-Hoover dynamics and Nos$\acute\textrm
  e$-Hoover chain dynamics}
In the case of Nos$\acute\textrm e$-Hoover dynamics~\cite{Nose,Hoover}
the heat bath is represented by an additional degree of freedom
$\zeta$ and effective mass $Q$ and 
 the equilibrium distribution function is the extended canonical
distribution~\cite{Hoover}
\begin{equation}
  \label{eq:nosehooverdist}
  \rho(p,q,\zeta)\propto e^{-\beta \left(H(p,q;\lambda)+\zeta^2/2Q\right)}
\end{equation}
where $H$ is the Hamiltonian of the system and $(q,p)$ are the
positions and momenta of the particles. Thus
relation~(\ref{eq:equilibrium}) is fulfilled if one identifies $x$
with $(p,q)$, $y$ with $\zeta$ and sets $\psi(y)\equiv \zeta^2/2Q$
and the Jarzynski identity holds.
In order to enhance ergodic sampling, the Nos$\acute\textrm e$-Hoover
approach is often augmented with chains of Nos$\acute\textrm e$-Hoover
thermostats~\cite{Martyna}. 
The corresponding equations of motion preserve the phase space
distribution
\begin{equation}
  \label{eq:nosehooverchaindist}
  \rho(p,q,\zeta_i)\propto e^{-\beta \left(H(p,q;\lambda)+\sum_{i=1}^M \zeta_i^2/2Q_i\right)}.
\end{equation}
that has the form of Eq.~(\ref{eq:equilibrium}) which is required for
the validity of the Jarzynski identity and the Crooks fluctuation
theorem.

\subsection{Gaussian-isokinetic dynamics}
In contrast to the Nos$\acute\textrm e$-Hoover dynamics that fixes the
temperature of the system the Gaussian isokinetic equations fix the
kinetic energy~\cite{isoEvans,isoHoover}.
In the thermodynamic limit the corresponding equilibrium
distribution~\cite{EvansMoriss} is the so called isokinetic
distribution
\begin{equation}
  \label{eq:isokineticdist}
  \rho(p,q,\xi)=\frac{e^{-\beta V(q;\lambda)}\delta(K(p)-K_0)}{\int dpdq e^{-\beta V(q;\lambda)}\delta(K(p)-K_0)}.
\end{equation}
This distribution is canonical in configuration space and
microcanonical in momentum space. If the kinetic energy of the system
is independent of the external parameter, i.e.
\begin{equation}
  \label{eq:hamiltonian}
  H(x;\lambda)=V(q;\lambda)+K(p),
\end{equation} 
the work performed on the system can be rewritten as
\begin{equation}
  \label{eq:WV}
    W=\int_0^{\tau}\mathrm{d}t\frac{\partial
    H(x(t),\lambda(t))}{\partial \lambda}\dot{\lambda}(t)=\int_0^{\tau}\mathrm{d}t\frac{\partial
    V(q(t),\lambda(t))}{\partial \lambda}\dot{\lambda}(t).
\end{equation}
Thus
\begin{eqnarray}
  \label{eq:delatWV}
    \Delta H&=&\int_0^{\tau}\mathrm{d}t\frac{\partial
    H(x(t),\lambda(t))}{\partial \lambda}\dot{\lambda}(t)+\int_0^{\tau} \mathrm{d}t\nabla_xH(x(t),\lambda(t))\cdot\dot{x}(t)\nonumber\\
&=&\Delta V=\int_0^{\tau}\mathrm{d}t\frac{\partial
    V(q(t),\lambda(t))}{\partial \lambda}\dot{\lambda}(t)+\int_0^{\tau} \mathrm{d}t\nabla_qV(q(t),\lambda(t))\cdot\dot{q}(t)
\end{eqnarray}
implying that the heat transfer to the system is given by
\begin{equation}
  \label{eq:QV}
  Q=\int_0^{\tau} \mathrm{d}t\nabla_xH(x(t),\lambda(t))\cdot\dot{x}(t)=\int_0^{\tau} \mathrm{d}t\nabla_qV(q(t),\lambda(t))\cdot\dot{q}(t).
\end{equation}
The equilibrium distribution~(\ref{eq:isokineticdist}) takes the form
of Eq.~(\ref{eq:equilibrium}) if one identifies $x\equiv q$, $y\equiv
p$, $\psi(y)\equiv\delta(K(p)-K_0)$ and replaces $H(x;\lambda)$ by
$V(q;\lambda)$. Since the definition of work and heat remain valid if
one replaces the Hamiltonian function by the potential energy, the
Jarzynski theorem and the Crooks fluctuation theorem are valid for
Gaussian isokinetic dynamics.

\section{Conclusion}
We have presented a proof of the Jarzynski theorem and the Crooks
fluctuation theorem. This proof is valid for systems evolving under
deterministic equations of motion that conserve an extended
factorizing canonical distribution of the form of
Eq.(\ref{eq:equilibrium}). Several dynamical systems that meet the
latter condition have been discussed. The crucial part of the
derivation is a relation between the heat transferred to the extended
system during the switching process and the phase space compression
factor of the dynamics which was established using Euler's expansion
formula.

\section{Acknowledgments}
We thank C. Jarzynski for useful discussions. This work was supported
by the Austrian Science Fund (FWF) under Grant No. P17178-N02.

\end{document}